\documentclass[pra,twocolumn,showpacs,preprintnumbers]{revtex4}

\usepackage{amsmath}
\usepackage{amsfonts}
\usepackage{amssymb}

\usepackage{graphicx}
\usepackage{dcolumn}
\usepackage{bm}

\usepackage{colordvi}
\usepackage{txfonts}

\begin{document}

\title{Automated and robust population transfer method for three-level system using oscillating dark states}

\author{Jun He}
\email{jhe@mail.ustc.edu.cn}

\author{Yong-Sheng Zhang}
\email{yshzhang@ustc.edu.cn}

\author{Xiang-Fa Zhou}
\author{Qun-Feng Chen}
\author{Guang-Can Guo}

\affiliation{Key Laboratory of Quantum Information, University of
Science and Technology of China(CAS), Hefei, Anhui 230026, China}

\date{\today}

\begin{abstract}
An automated and robust method for adiabatic population transfer and
the preparation of an arbitrary quantum superposition state in
atomic system using the oscillating dark states (ODS) is presented.
Quantum state of a three-level $\Lambda$ configuration atomic system
oscillates periodically between two ground levels, when two pairs of
classical detuning laser fields driving the system into the ODS
under evolving adiabatic conditions. The decoherence of the ODS
evolution is greatly suppressed, and the oscillation is very stable,
therefore adiabatic population transfer and the preparation of an
arbitrary quantum superposition state of atomic system can be
completed accurately and conveniently.

\end{abstract}

\pacs{42.50.Gy, 32.80.Qk}

\maketitle

\section{INTRODUCTION}

Dark state or coherent population trapping (CPT), which is null
eigenvalue state of the interaction Hamiltonian, has given rise to
growing interests for past few decades, since its first discovery in
1976~\cite{alzetta1976}. If a three-level $\Lambda$ atomic system is
coherently trapped in dark states, there is no absorption of
incident laser fields even in the presence of resonant transitions,
due to destructive quantum interference between two transition
pathways. Well-known dark state or
CPT~\cite{Arimondo-Orriols,gray,Arimondo,EIT-RMP-2005,CPT-RMP-1998,Lukin1999,quantum-optics,dark-state-polariton2000}
involves extensive applications in many fields such as
electromagnetically induced transparency (EIT)~\cite{harris1997},
lasing without inversion~\cite{lasing1,lasing2,lasing3}, the
enhancement of the refractive index~\cite{enhancement of the
refractive index 1,enhancement of the refractive index 2}, adiabatic
population transfer~\cite{oreg1984,bergmann1989}, subrecoil laser
cooling~\cite{cohen1994}, and atom interferometry~\cite{chu}.
Coherent population transfer using stimulated Raman adiabatic
passage(STIRAP)
technique~\cite{CPT-RMP-1998,annu_rev_phys_chem2001,bergmann1989,Gaubatz1990},
make it robust and efficient to prepare atoms and molecules in a
well defined required state in many fields such as spectroscopy,
collision dynamics, and atomic physics. The Stokes and pump laser
fields in a kind of counterintuitive order under two-photon
resonance and adiabatic evolution carry out complete coherent
population transfer, and the STIRAP technique is insensitive to
variations of laser pulse shape, intensity, and laser frequency. It
is also of crucial importance that a required coherent superposition
state is prepared for atoms and molecules. Fractional STIRAP
technique~\cite{Marte1991,J_Phys_B1999}, where the Stokes laser
arrives before the pump laser and both terminate simultaneously
maintaining a constant ratio of relative amplitudes, can create an
arbitrary required coherent superposition state. The technique is
also insensitive to pulse delay, intensity and frequency of laser
fields. Another adiabatic transfer~\cite{Unanyan1998} using a tripod
linkage by three laser fields in a four-level system can also to
obtain good result.

Here we propose an automated method for adiabatic population
transfer and the preparation of an arbitrary quantum superposition
state in a three-level $\Lambda$ configuration atomic system, using
the oscillating dark states (ODS), where probability amplitudes of
the system state oscillate periodically between two ground levels.
The state to be prepared can be retrieved at a predictable time, and
this method is insensitive to the initial state.

The paper is organized as follows. The next section contains the
discussion of the oscillating dark states (ODS) and evolving
adiabatic conditions, under which the system follows the stable ODS
evolution. In Sec. III, we introduce the master equation for a
practical incoherent atomic system, and simulate the ODS evolution,
which then be used for adiabatic population transfer and the
preparation of an arbitrary quantum superposition state. We give
some discussion and summarize our conclusions in Sec. IV.

\section{OSCILLATING DARK STATES}

\begin{figure}
\centering
\includegraphics[scale=0.20]{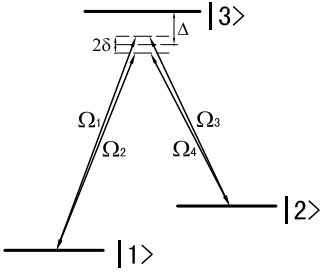}%
\caption{\label{final-big2} Two pairs of classical laser fields
$\Omega_1$ , $\Omega_2$ and $\Omega_3$ , $\Omega_4$ couple two lower
ground levels $|1\rangle$ , $|2\rangle$ to a single upper level
$|3\rangle$ , respectively, with their corresponding detunings
$\Delta_1$ , $\Delta_2$ and $\Delta_3$ , $\Delta_4$, and their fixed
phases: $\phi_1$=$\phi_2$+$\pi$, $\phi_3$=$\phi_4$, in a three-level
$\Lambda$ configuration atomic system.}
\end{figure}

The three-level $\Lambda$ configuration atomic system is shown in
Fig.~\ref{final-big2}. The ODS is implemented with two pairs of
classical detuning laser fields coupling two lower ground levels
$|1\rangle$ , $|2\rangle$ to a single upper level $|3\rangle$,
respectively. We assume that the $|1\rangle$-$|2\rangle$ transition
is always dipole-forbidden. The time-dependent interaction
Hamiltonian in the interaction picture that describes the atom-laser
coupling within the dipole and rotating wave approximation (RWA) in
the rotating frame reads ($\hbar$=$1$),
\begin{equation}
H_{int}=\left[
\begin{array}{c}~0~~~~~0~~~~P^*\\~0~~~~~\Delta'~~Q^*\\P~~~~Q~~~~\Delta
\end{array} \right]\;, \label{eq:interaction}
\end{equation}
where
$P$=$-ie^{-i\phi_{12}}\Omega_{12}\sin(\frac{\Delta_1-\Delta_2}{2}t)$,
$Q$=$-e^{-i\phi_{34}}\Omega_{34}\cos(\frac{\Delta_3-\Delta_4}{2}t)$,
$\Delta$=$\frac{1}{2}(\Delta_1+\Delta_2)$, and
$\Delta'$=$\frac{1}{2}[(\Delta_1+\Delta_2)-(\Delta_3+\Delta_4)]$.
$\Delta_i$=$\omega_{31}$-$\omega_i$($i$=$1,2$), and
$\Delta_j$=$\omega_{32}$-$\omega_j$($j$=$3,4$) are the detunings of
two pairs of laser angular frequencies $\omega_i$($i$=$1,2$) and
$\omega_j$($j$=$3,4$) from the corresponding atomic transitions
$\omega_{31}$ and $\omega_{32}$. Real $\Omega_i$ and $\phi_i$
($i$=$1,2 ; 3,4$) are corresponding Rabi frequencies and phases of
two pairs of laser fields, and we assume that
$\Omega_1$=$\Omega_2$=$\Omega_{12}$,
$\Omega_3$=$\Omega_4$=$\Omega_{34}$,
$\phi_1$=$\phi_2$+$\pi$=$\phi_{12}$, and
$\phi_3$=$\phi_4$=$\phi_{34}$, with the difference
$\Delta\phi$=$\phi_{12}$-$\phi_{34}$.
The eigenstates of the interaction
Hamiltonian~(\ref{eq:interaction}), which can be generally described
in terms of the \textquotedblleft mixing angles\textquotedblright
$\theta$ and $\varphi$ , are given as
\begin{equation}
|a_+\rangle=-ie^{i\Delta\phi}\sin\theta\sin\varphi|1\rangle+\cos\theta\sin\varphi|2\rangle-\cos\varphi
e^{-i\phi_{34}}|3\rangle\;, \label{eq:a+}
\end{equation}
\begin{equation}
|a_0\rangle=\cos\theta|1\rangle-i\sin\theta
e^{-i\Delta\phi}|2\rangle\;,
%\fleqn
\label{eq:a0}
\end{equation}
\begin{equation}
|a_-\rangle=-ie^{i\Delta\phi}\sin\theta\cos\varphi|1\rangle+\cos\theta\cos\varphi|2\rangle+\sin\varphi
e^{-i\phi_{34}}|3\rangle\;, \label{eq:a-}
\end{equation}
where the mixing angles that dependent upon time $t$ and Rabi
frequencies are written as
\begin{equation}
\tan\theta=\frac{\Omega_{12}\sin(\frac{\Delta_1-\Delta_2}{2}t)}
{\Omega_{34}\cos(\frac{\Delta_3-\Delta_4}{2}t)}\;, \label{eq:tan}
\end{equation}
\begin{equation}
\tan2\varphi=\frac{2\sqrt{|P|^2+|Q|^2}}{\Delta}\;,
\end{equation}
when we meet the condition $\Delta'=0$, i.e.
$(\Delta_1+\Delta_2)$=$(\Delta_3+\Delta_4)$. The eigenstate
$|a_0\rangle$ has no contribution from $|3\rangle$, and this ODS
corresponds the null eigenvalue of interaction
Hamiltonian~(\ref{eq:interaction}). If a system is in state
$|a_0\rangle$, there is no possibility of excitation to $|3\rangle$
and subsequent spontaneous emission, thus the atom is effectively
decoupled from laser fields. The eigenvalues of the pair of states
$|a_\pm\rangle$ which contain a component of all three bare atomic
states, are shifted up and down by an amount $\lambda_\pm$,
\begin{equation}
\lambda_\pm=\frac{1}{2}[\Delta\pm \sqrt{\Delta^2+4|P|^2+4|Q|^2} ]\;.
\label{eq:lambda-pm}
\end{equation}
The probability amplitudes of the system state can oscillate
periodically between two ground levels $|1\rangle$ and $|2\rangle$
upon time $t$, when the system is driven into the ODS by laser
fields with relatively fixed detunings and phases of four laser
fields.

\par Evolving adiabatic conditions, under which the system avoids
diabatic coupling to bright states and is always in some dark states
if it is initially in a certain one, is considered, since the ODS
follows the evolution of a time-dependent Hamiltonian
(\ref{eq:interaction}). The change rate of the mixing angle $\theta$
must be small compared with the separation of the corresponding
eigenvalues~\cite{Messiah1962,Gaubatz1990},
\begin{equation}
|\frac{\textit{d}}{\textit{dt}}\theta|^2\ll|\lambda_0-\lambda_\pm|^2\;.
\label{eq:totalconditon}
\end{equation}
Let $\Omega_{12}$=$\Omega_{34}$=$\Omega$ (fixing Rabi frequencies
when the system evolves) and
$\Delta_1$-$\Delta_2$=$\Delta_3$-$\Delta_4$=2$\delta$, thus we can
obtain evolving adiabatic conditions according to
(\ref{eq:totalconditon}) as
\begin{equation} |\delta|^2\ll \frac{1}{4}|\Delta\pm
\sqrt{\Delta^2+4\Omega^2}|^2\;, \label{eq:Evolving adiabatic
conditions}
\end{equation}
which changes into $|\delta|\ll$$\Omega$, when $\Delta=0$, i.e.
$\Delta_1$=-$\Delta_2$ and $\Delta_3$=-$\Delta_4$. Evolving
adiabatic conditions for the ODS ensure that the values of all
parameters can be established in advance, because the evolution
process is automated by laser-atom system itself, in contrast to
ordinary dark states~\cite{quantum-optics,bergmann1989}, where we
manually with experimental instruments change relative values of two
Rabi frequencies to attain different dark states and to utilize
STIRAP technique.

\section{ADIABATIC POPULATION TRANSFER AND QUANTUM STATE PREPARATION}

A process for adiabatic population transfer and quantum state
preparation is as follows: first, the initial instantaneous
eigenstate of the whole system, which is determined by chosen
interaction Hamiltonian, must be the atomic initial state, and here
we assume that the atomic initial state is $|1\rangle$. Then we
concurrently and rapidly upload two pairs of laser fields as shown
in Fig.~\ref{final-big2}, and the system periodically evolves
subsequently. Rapid unloading all laser fields simultaneously
completes a perfect population transfer from $|1\rangle$ to
$|2\rangle$ after ($\frac{1}{4}$+$\frac{n}{2}$) ($n$=$0,1,2,...$,
integral number) oscillation periods($T$=$2\pi/\delta$) if we want
to make an adiabatic population transfer, and retrieves a required
quantum superposition state after time of ($t_0$+$nT$)($t_0$ is
evolution time when the system firstly evolves from $|1\rangle$ to
our required state, which can be easily predicted from (\ref{eq:a0})
where a fixed $\Delta\phi$ should be calculated in advance) if we
use the same experimental setup for the preparation of an arbitrary
quantum superposition state.

When we upload/unload two pairs of laser fields, uploading/unloading
adiabatic conditions are necessary besides evolving adiabatic
conditions~(\ref{eq:Evolving adiabatic conditions}), in order to
keep the least diabatic coupling between bright states and dark
states of the system. We assume a concurrent increase/decrease of
four laser fields in a short time $\tau$($\tau$$\ll$$T$), i.e.
$\Omega_{1}$=$\Omega_{2}$=$\Omega_{3}$=$\Omega_{4}$=$\Omega$ and
$\frac{\textit{d}}{\textit{dt}}$$\Omega_{1}$=
$\frac{\textit{d}}{\textit{dt}}$$\Omega_{2}$=
$\frac{\textit{d}}{\textit{dt}}$$\Omega_{3}$=
$\frac{\textit{d}}{\textit{dt}}$$\Omega_{4}$. Thus, we can obtain
the same result as evolving adiabatic conditions in
(\ref{eq:Evolving adiabatic conditions}) for uploading/unloading
adiabatic conditions, according to (\ref{eq:totalconditon}).

Time-dependent interaction Hamiltonian in interaction picture that
describes atom-laser coupling can be written as
\begin{eqnarray}
H'_{int}=-\frac{1}{2}[\Omega_{1}e^{-i\phi_1}e^{i\Delta_1t}\sigma_{31}+\Omega_{2}e^{-i\phi_2}e^{i\Delta_2t}\sigma_{31}\nonumber\\
+\Omega_{3}e^{-i\phi_3}e^{i\Delta_3t}\sigma_{32}+\Omega_{4}e^{-i\phi_4}e^{i\Delta_4t}\sigma_{32}+H.c.]\;,
\label{eq:H-int'}
\end{eqnarray}
where $\sigma_{ij}$=$|i\rangle\langle j|$ ($i,j$=$1,2,3$) are atomic
projection operators. A master
equation~\cite{EIT-RMP-2005,cohen1992}, using density matrix $\rho$,
\begin{eqnarray}
\frac{\textit{d}\rho}{\textit{dt}}=-i[H'_{int},\rho]+\frac{\Gamma_{31}}{2}[2\sigma_{13}\rho\sigma_{31}-\sigma_{33}\rho-\rho\sigma_{33}]\nonumber\\
+\frac{\Gamma_{32}}{2}[2\sigma_{23}\rho\sigma_{32}-\sigma_{33}\rho-\rho\sigma_{33}]\nonumber\\
+\frac{\gamma_{3\textit{deph}}}{2}[2\sigma_{33}\rho\sigma_{33}-\sigma_{33}\rho-\rho\sigma_{33}]\nonumber\\
+\frac{\gamma_{2\textit{deph}}}{2}[2\sigma_{22}\rho\sigma_{22}-\sigma_{22}\rho-\rho\sigma_{22}]\nonumber\\
+\frac{\Gamma_{21}}{2}[2\sigma_{12}\rho\sigma_{21}-\sigma_{22}\rho-\rho\sigma_{22}]\;,
\label{eq:master-equation}
\end{eqnarray}
is required, because the decoherence of a practical atomic system
must be considered. $\Gamma_{31}$ and $\Gamma_{32}$ represent rates
of spontaneous emission from level $|3\rangle$ to $|1\rangle$ and
$|2\rangle$, respectively, and rates $\gamma_{2\textit{deph}}$ and
$\gamma_{3\textit{deph}}$ describe energy-conserving dephasing
processes. $\Gamma_{21}$ is longitudinal relaxation from level
$|2\rangle$ to $|1\rangle$, and it is much small compared with
$\gamma_{2\textit{deph}}$. We can define coherence decay rates as
$\gamma_{31}$=$\Gamma_{31}$+$\Gamma_{32}$+$\gamma_{3\textit{deph}}$,
$\gamma_{21}$=$\Gamma_{21}$+$\gamma_{2\textit{deph}}$, and estimate
$\Gamma_{21}$ at $1/10$ of $\gamma_{2\textit{deph}}$ in the
following analysis.

\begin{figure}
\includegraphics[scale=0.5]{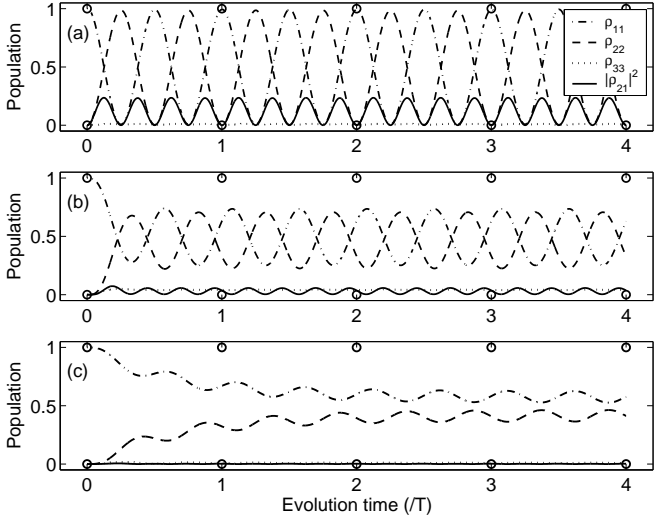}
\caption{\label{fig:2} Probability amplitudes evolution of initial
state $|1\rangle$ finds expression in the time-dependent populations
of three levels $\rho_{11}$, $\rho_{22}$, $\rho_{33}$, and coherence
term between two ground levels $|\rho_{21}|^2$ with
$\Delta_1=\Delta_3=0.3$, $\Delta_2=\Delta_4=0.2$,
$\gamma_{2\textit{deph}}=0.02$, $\Gamma_{21}=0.002$, for different
case: (a). $\Omega=2$, (b). $\Omega=0.2$, (c). $\Omega=0.08$,
respectively. Randomly chosen $\Delta \phi$ (adopted for different
required superposition state, but having no effect on coherent
population transfer process) gives the same periodical oscillation.
All parameters are in the units of $\gamma_{31}$, and total
evolution time $t$=4$T$ is marked.}
\end{figure}

The evolution of the ODS is simulated numerically as shown in
Fig.~\ref{fig:2}, when the initial system is in $|1\rangle$ and a
fixed $\Delta\phi$ should be calculated in advance (for a required
arbitrary final state), and probability amplitudes evolution of the
ODS with different Rabi frequencies $\Omega$ is analyzed. The
populations of two ground lower levels, $\rho_{11}$ and $\rho_{22}$,
oscillate periodically after we applying interaction Hamiltonian
(\ref{eq:H-int'}), and the population of upper level $\rho_{33}$
stays almost zero in total evolution stage, when evolving adiabatic
conditions~(\ref{eq:Evolving adiabatic conditions}) are well
satisfied for the case in Fig.~\ref{fig:2}(a). The same periodical
oscillation of the coherence term $|\rho_{21}|^2$ between two ground
levels as $\rho_{11}$ and $\rho_{22}$, and
$|\rho_{21}|^2$$\approx$$\rho_{11}$$\cdot$$\rho_{22}$, imply a
coherent evolution of quantum superposition states, i.e. the ODS. In
case (b), Rabi frequency $\Omega$ is decreased, and evolving
adiabatic conditions are not well met, which results in a slight
occupation of population $\rho_{33}$, but the oscillations of
population $\rho_{11}$, $\rho_{22}$ and of coherence term
$|\rho_{21}|^2$ still dominate the whole evolution process. When we
decrease $\Omega$ further, laser fields are too weak to make upper
level $|3\rangle$ occupied despite the violation of evolving
adiabatic conditions, and the coherence term of density matrix
$|\rho_{21}|^2$ keeps $0$ all the time in case (c), which implies
there is no coherent evolution of the system state.

From case (c) to (a), the value of Rabi frequencies $\Omega$ of
laser fields increased little by little, it can be seen that
coherent reconstruction mechanism of laser fields becomes superior
gradually to the decoherence effect of the system in case (c) and
(b), and entirely dominates periodical oscillation of the system in
case (a). A complete population transfer between two lower ground
levels $|1\rangle$ and $|2\rangle$, and the preparation of an
arbitrary quantum superposition state in an automatical manner,
using a coherent evolution process of the ODS in case (a), is
reliable. Here upper level $|3\rangle$ population is much little and
the oscillation of probability amplitudes of two ground levels
$|1\rangle$ and $|2\rangle$ is periodical, since the decoherence
between two ground levels $|1\rangle$ and $|2\rangle$ can be greatly
suppressed, owing to coherent reconstruction contribution from laser
fields. It can be seen that there are many optional retrieval time,
with retrieval interval $T/2$ for an adiabatic population transfer
process and with interval $T$ for the preparation of an arbitrary
quantum superposition state.

\begin{figure} %height=4cm,width=6cm
\includegraphics[scale=0.34]{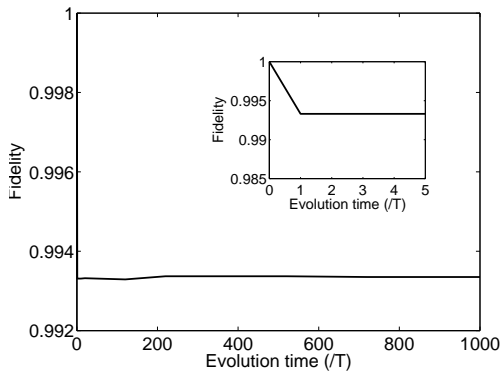}
\caption{\label{fig:t_VS_fidelity} Fidelity of initial state
$|1\rangle$ and retrieved states of the system after different
evolution time of the system (only integral period evolution
observed, and up to a maximum $1000T$ here), with $\Omega=2$,
$\Delta_1=\Delta_3=0.3$, $\Delta_2=\Delta_4=0.2$,
$\gamma_{21}=0.022$. After the first one $T$ (see inset), High
fidelity above $0.95$ is available and keep stable at all times,
even if $\gamma_{21}\cdot t$$\gg$1, where different initial states
of system almost give the same result.} %, and independent variable $t$ whatever the initial state of
%atomic system is, besides $|1\rangle$ here.
\end{figure}

With the contribution from applied laser fields, the state of the
system evolves periodically and stably despite decay factor
$\gamma_{21}\cdot t$ expressed in coherence term of the system
density matrix $|\rho_{21}|^2$. When two pairs of laser fields are
applied, we can also consider the long time behavior of state
evolution of the system and analyze the fidelity of initial state
$|1\rangle$ and retrieved states of the system after different
evolution time $t$ in Fig.~\ref{fig:t_VS_fidelity}, where only
integral period evolution ($t$=$nT$) is observed. $\gamma_{21}$ is
fixed at $0.022$ in the units of $\gamma_{31}$ (reference to $D_1$
line of $^{87}Rubidium$ with
$\gamma_{31}$$\sim$$36.10\times10^6$$s^{-1}$ and estimated
$\gamma_{2\textit{deph}}$$=$$0.02$$\gamma_{31}$), and different
total evolution times $t$ is given. It can be seen that, when total
evolution time $t$ is up to long enough $1000$$T$, where
$\gamma_{21}\cdot t$$>$$2000$, high fidelity above $0.95$ is
surprisingly available, which implies almost a coherence preserved
and pure periodical evolution of the system. Fidelity $\textit{F}$
decreases slightly in the first one $T$, during which a certain
complicated amplitude decay of the ODS experiences, and stays
unchangeably after it, which also can be found in the very beginning
of evolution process in Fig.~\ref{fig:2}. High fidelity $\textit{F}$
mainly depends on relatively large value of $\Omega$ and always can
be obtained by controlling laser fields.

\section{DISCUSSION AND CONCLUSION}

The stability of frequency difference $2\delta$, $\pi$ phase
difference, and $\Delta\phi$ of each pair of laser fields determine
mainly the precision of the ODS evolution. The population transfer
efficiency and an ideal required superposition state mainly depend
on the exact retrieval time that can be easily calculated according
to~(\ref{eq:a0}). Two pairs of laser fields should be
uploaded/unloaded spontaneously instead of counterintuitive and
delayed uploading of one pair Stokes and pump laser fields. A
concurrent uploading/unloading time $\tau$$\sim$$10^{-8}$$s$ (a
reasonable $0.01T$ in our analysis which is short enough to avoid a
certain decay of $\rho$ at this stage) can be carried out using an
optical switch, provided all the parameters are chose as ours, which
can be tuned to change the oscillation period $T$. Laser fields may
not be monochromatic, but the stability of relative phases of two
pairs of laser fields, which can be easily achieved by
acousto-optical modulation(AOM) device, can also guarantee almost
the same result, despite a slow collective phase drift. A chosen
$\delta$ whose value is larger than the linewidth of laser fields
can avoid undesirable interference of laser fields and optical
pumping. The ODS method is immune to Doppler frequency shift, pulse
shape of laser fields, and concurrent fluctuation of Rabi
frequencies $\Omega_i$ ($i$=$1,2 ; 3,4$). Well developed laser
techniques can meet all above requirements to obtain a perfect
adiabatic population transfer and the preparation of an arbitrary
quantum superposition state in atomic system. The greatly suppressed
decoherence of the system according to our numerical calculation
suggests the stability of the ODS evolution. Numerous optional
retrieval time owing to the property of the ODS evolution is
available, with retrieval interval $T/2$ for an adiabatic population
transfer process and with interval $T$ for the preparation of an
arbitrary quantum superposition state, respectively. The transfer
process is insensitive to the initial state of the atomic
system~\cite{ottaviani-simon}, i.e., we can carry out complete
coherent population transfer and obtain a required state at right
retrieval time based on precise ODS evolution of atom-laser system,
whatever the initial state of atomic system is, which is totally
different from the STIRAP and fractional STIRAP technique. Different
initial states of the system besides $|1\rangle$ here, to which the
corresponding initial Hamiltonian of the system can be chosen to
match the initial instantaneous eigenstate, give almost the same
good result. Optical pumping is not necessary any longer in contrast
to conventional STIRAP technique, since it is not essential that
initial state of the system must be any single lower state
$|1\rangle$ or $|2\rangle$ when the automated ODS method is adopted.
The ODS can be implemented in a three-level atomic system by using
at least three laser fields in an appropriate configuration, but we
use two pairs of laser fields in a symmetric scheme, in order to
obtain complete oscillation of probability amplitudes between two
ground levels. The ODS is similar to the famous Rabi resonance in a
two-level atomic system driven by one resonant laser
field~\cite{quantum-optics}, where we can not find the same
preserved coherence.

In conclusion, we propose an automated method using the ODS for a
perfect adiabatic population transfer and the preparation of an
arbitrary quantum superposition state in atomic system. The
preserved coherence, the automated manner, numerous optional
retrieval time, an arbitrary initial state of system, make the ODS a
potential method for physics process in atomic physics.

\begin{center}
\textbf{ACKNOWLEDGMENTS}
\end{center}

We thank Carlo Ottaviani and Christoph Simon for helpful discussion.
This work was funded by the National Fundamental Research Program
(Grant No. 2006CB921907), the National Natural Science Foundation of
China (Grant No. 10674127 and No. 60621064), the Innovation Funds
from the Chinese Academy of Sciences, Program for NCET and
International cooperation program from CAS.

\end{document}